\documentclass[12pt]{article}

\usepackage{amsmath,amsfonts,amssymb,latexsym,stmaryrd,mathrsfs}

\newtheorem{theorem}{Theorem}[section]

\numberwithin{equation}{section}
\def\be{\begin{equation}}
\def\ee{\end{equation}}
\def\bq{\begin{eqnarray}}
\def\eq{\end{eqnarray}}
\def\beq{\begin{eqnarray*}}
\def\eeq{\end{eqnarray*}}

\begin{document}
\begin{titlepage}
\begin{flushright}
{\tt CERN-PH-TH/2014-176}
\end{flushright}


\begin{center}
{\Huge Ambient cosmology and spacetime singularities}

\vspace{1cm}
{\large Ignatios Antoniadis$^{1,2*}$, Spiros Cotsakis$^{3,4,5,\dagger}$ }\\

\vspace{0.5cm}

$^1$ {\normalsize Albert Einstein Center for Fundamental Physics, Institute for Theoretical Physics}\\
{\normalsize Bern University, Sidlerstrasse 5 CH-3012 Bern, Switzerland}\\

$^2$ {\normalsize Ecole Polytechnique, F-91128 Palaiseau, France}\\

$^3$ {\normalsize  Department of Physics, CERN - Theory Division}\\
{\normalsize CH--1211 Geneva 23, Switzerland}\\

$^4$ {\normalsize School of Applied Mathematics and Physical Sciences \\
National Technical University, 15780 Athens, Greece} \\

\vspace{2mm} {\normalsize {\em E-mails:}
$^*$\texttt{ignatios.antoniadis@cpht.polytechnique.fr},
$^\dagger$\texttt{spiros.cotsakis@cern.ch}
}
\end{center}


\begin{abstract}
\noindent We present a new approach to the issues of spacetime singularities and cosmic censorship in general relativity. This is based on the idea that standard 4-dimensional spacetime is the conformal infinity of an  ambient metric for the 5-dimensional Einstein equations with fluid sources. We then  find that the existence of spacetime  singularities in four dimensions is constrained by asymptotic properties  of the ambient 5-metric, while the non-degeneracy of the latter crucially depends on cosmic censorship holding on the boundary.

\end{abstract}

\begin{center}
{\line(5,0){280}}
\end{center}

$^5${\small On leave from the University of the Aegean, 83200 Samos, Greece.}

\end{titlepage}
\section{Introduction}
The geometry and physics of braneworlds \cite{hw,add,aahd,bin} and holographic ideas \cite{t'} such as  the AdS/CFT correspondence \cite{malda,wit} have provided conclusive evidence that  new structures may exist in higher dimensions described by some metric $g_5$, with the four-dimensional, general relativistic world confined in a suitable subspace $g_4$. The precise nature of the theory describing our universe that forms the basis of this is likely a version of string theory whose realization is presently unknown, and the same is true in such  frameworks of the possible resolution of the well-known issues of Einstein's theory, such as the existence and nature of spacetime singularities and the justification of cosmic censorship.

Concerning the problem of singularities \cite{p64,hp} and its possible resolution in higher dimensions, we note that the issue is further complicated by the appearance of new singularities in the geometry of the extra dimensions  and the associated problems that this brings about, cf.  \cite{rs1}-\cite{Forste2} (and Refs. therein). Also, cosmic censorship \cite{p99} seems to emerge as an inherent property of four-dimensional general relativistic metrics $g_4$, unconnected to the higher-dimensional structures presumably responsible to the relativistic singularity resolution.

In this work, we view the four-dimensional relativistic world as a suitable asymptotic and holographic limit of structures that may exist only in higher dimensions. We know \cite{skot} how to asymptotically split a five-dimensional space $g_5$ containing a `braneworld' \cite{ack} $g_4$, and starting with our asymptotic splitting solutions in the form $(M\times \mathbb{R},g_5=a^2(y)g_4+dy^2)$, in the first part of this paper we introduce a new way to connect the properties of the four-dimensional general relativistic world  to a possible higher-dimensional theory (in 5 dimensions).

The basic idea of this work is that our world is the conformal infinity of a certain `ambient' metric in one higher dimension. (We use the adjective `ambient' to declare an analogy with the Fefferman-Graham ambient construction \cite{fef}.) The ambient metric $(V=M\times \mathbb{R},g_+)$ satisfies the 5-dimensional Einstein equations with fluid sources, but is constructed  locally in an open neighborhood of the boundary, 4-dimensional conformal geometry $(M,[g_4])$, and has a suitable metric  $\mathring{g}|_M$ in the conformal class as its conformal infinity. Because of its dependence on formal power series, this construction relies on both asymptotic as well as  holographic properties of the higher-dimensional space, and possesses various novel properties. However, our approach is distinct from the general braneworld approach as well as the AdS/CFT and related approaches, in that in the former the brane cannot be the conformal infinity of the bulk geometry without spoiling the solutions completely, whereas in the latter one does not deal with gravitational effects on the boundary.

This paper falls into three parts. The first part, up to Section 3,  deals with the general motivation and outlines the basic features of our approach. In the next section we discuss some of the elements  of standard braneworld cosmology relevant to the present context, and in  Section 3 we outline in a series of distinct steps an alternative approach and method we use to deal with several of the features of the standard braneworld model. In Part 2, which includes Sections 4-6, we expand on the various justifications of the steps involved in our construction and describe the normal form of the ambient metric, its conformal infinity and the asymptotic conditions, which are all key elements of ambient cosmology. In the third part of this work, Sections 7, 8, we discuss implications of this construction for the existence of spacetime singularities and the question of cosmic censorship in the 4-dimensional boundary of the ambient world. 

\section{The standard braneworld cosmology}
In previous works \cite{ack}, we studied the asymptotic properties of bulk 5-geometries $(V,g_5)$ containing an embedded 4-dimensional braneworld $(M,g_4)$ that was either a 4-dimensional  Minkowski, or de Sitter, or Anti-de Sitter spacetime, and showed that, in general, asymptotic solutions have a form dictated by the method of asymptotic splittings \cite{skot}, namely,
\be\label{scale}
a(y)=y^p\sum_{i=0}^{\infty}c_iy^{i/s},\quad y\rightarrow 0,
\ee
where the first constant $c_0$ in the series is nonzero (this is the dominant balance).
Here, $y$ denotes the coordinate of the extra dimension in the 5-dimensional geometry $g_5$, where
\be\label{metric1}
g_5=a^2(y)g_4+dy^2,
\ee
 $g_4$ being a 4-dimensional braneworld metric with signature $(-+++)$, and $a(y)$ is  a warp factor in the $g_5$ geometry. The metric $g_5$ is taken to satisfy the 5-dimensional Einstein equations
\be\label{ein}
G_{AB}=T_{AB},
\ee
where $A,B=1,2,3,4,5$, and $T_{AB}$ is the stress tensor of an analog of a perfect fluid with equation of state $P=\gamma \rho$, where the `pressure' $P(y)$ and the `energy density' $\rho(y)$ depend only on $y$,\footnote{In the following, we drop the quotes for notational simplicity, even if $T_{AB}$ is not a perfect fluid in the usual cosmological sense.} filling the 5-dimensional  geometry (other cases like a bulk scalar field or a mixture of bulk fluids  considered in \cite{ack} also lead to the general form (\ref{scale}) for the warp factor of the 5-dimensional geometry (\ref{metric1})). These equations become, for the metric (\ref{metric1}), a dynamical system of the form $\dot{x}=f(x)$, with the solution vector having the form  $x=(a,\dot{a},\rho)$, and $f$ being a suitable, smooth field (cf. the references mentioned above). The asymptotic solutions for the energy density $\rho$ also have a form similar to (\ref{scale}).

Further, the constants $c_i,i=0,1,2,\dots$ in (\ref{scale})  are determined recursively by the method of asymptotic splittings starting at 0-th order with the dominant balance form $c_0y^p,p=m/l\in\mathbb{Q},m\in\mathbb{Z}, l\in\mathbb{N}$ (with the dominant balance constant $c_0$   nonzero), and proceeding in a term-by-term fashion, while $s$ is defined to be any common multiple of the denominators of
the positive eigenvalues of the Kovalevskaya matrix (cf. \cite{skot}).

This procedure leads to asymptotic solutions given generically by (\ref{scale}) (and similarly for the other unknowns), that is  Puiseux or Fuchsian series (meaning series with fractional exponents, and with or without a constant first term  respectively) describing the geometry locally in a small neighborhood around the location of the brane at $y=0$, or at infinity.

The braneworld setup described above has the following important properties: In the presence of a non-trivial fluid
\begin{itemize}
\item all 5-dimensional solutions are singular at a finite, arbitrary distance from the position of the brane, and the metric $g_5$ cannot be continued to arbitrary values in the $y$-dimension
    \item the properties of the metric $g_4$ do not follow from those of the bulk metric $g_5$
\item there is no conformal infinity for the 5-dimensional geometry.
\end{itemize}
These properties are common not only for the models in \cite{ack} but in fact, they are characteristic for all models of references \cite{rs1}-\cite{Forste2} (and related references therein). The first point means that there are genuine singularities in the metric and the curvatures in the 5-dimensional geometry which do not allow the metric to be continued smoothly beyond them.

For the second point, we note that  in most models of this sort the 4-metric is either taken to be fixed (e.g., Minkowski), or satisfying a 4-dimensional version of the Einstein equations with induced matter fields on the braneworld. In the latter case, although there are important differences, singularities are a typical feature of the evolution, in much the same way as in standard general relativity.

For the third point in the enumeration above, we note that in braneworld models the brane represents some kind of `boundary' for the 5-dimensional spacetime, often a domain wall with suitable boundary conditions, but it can never be a \emph{conformal} boundary. The reason is very simple: Trying to bring a solution of the form (\ref{metric1}) for the  warp factor of the 5-dimensional geometry (having all the required properties analyzed in detail in the references given above) to a suitable `conformal infinity' form $\tilde{g}_5=\Omega^2g_5$ by multiplying it by a conformal factor,  would completely destroy the properties of the original solution.
A consequence of this is that there is no possibility of a holographic interpretation and no way to realize a boundary CFT.

A closely related issue is that an AdS/CFT or related approach (such as the so-called `holographic renormalization cf. e.g., \cite{ske02}) could not really be used for the problems we have in mind here, simply  because there is no gravity on the boundary. In such approaches, one starts from  solutions in the bulk satisfying Einstein's equations with sources there, and then  produces a conformal structure on the boundary on which a CFT resides without any mention of possible \emph{gravitational} effects on the boundary.

\section{The ambient cosmology}
In this paper, we follow an inverse route and present a new approach that is distinctly different from \emph{any} AdS/CFT-based or braneworld approach. Starting  from a \emph{given} metric defined on the boundary, we then consider the conformal structure of that boundary metric in the sense of \cite{pen86}, and use that to construct in a series of steps, a 5-dimensional metric with the property that after the construction it returns a suitable 4-metric belonging to the conformal structure on the boundary that we started with. In this sense, our approach is closer in spirit to the original Fefferman-Graham construction.

In the resulting `ambient', 4-dimensional universe introduced here,  the classical singularities and the question of cosmic censorship in the 4-dimensional conformal boundary of the 5-dimensional ambient space acquire new meanings. The general procedure consists of starting with a conformal structure on a 4-dimensional boundary, and constructing the ambient 5-metric corresponding to that structure which satisfies the 5-dimensional Einstein-fluid equations and has a nice  conformal infinity.

In particular, we prove that corresponding to each 4-metric $g$ in the conformal class $[g_4]$ of another, well-behaved, 4-metric $g_4$ on the 4-boundary $M$, that is  $g_4=\Omega^2g$, there is a new 4-metric $\mathring{g}|_M$, a constant rescaling of $g_4$, which is the conformal infinity of the ambient 5-metric $g_{+}$ of the metric $g$. The ambient metric $g_{+}$ is defined on the `ambient' spacetime $V=M\times\mathbb{R}$ and satisfies the 5-dimensional Einstein equations with a fluid source.

More specifically, we show how to extend the 5-metric (\ref{metric1}) corresponding to the `nice' 4-metric $g_4$, to a new, 5-dimensional  metric $g_{4,+}$ (this is the ambient metric) on $V$ (a point in $V$  has coordinates $(x^\mu,w)$), such that the latter has a nice conformal infinity. Then we show that any 4-metric $g$ on the boundary $M$ that belongs to the conformal class of $g_4$ has itself an ambient 5-metric $g_+$ with conformal infinity $(M,\mathring{g}|_M)$ which is a constant rescaling of the nice metric $g_4$. Hence, our construction produces from a given metric $g\in[g_4]$ its ambient metric whose conformal infinity $\mathring{g}|_M$ acquires various important and improved  properties over the original 4-metric $g$.

The aforementioned extension is achieved in a series of steps as follows:
\begin{enumerate}
\item Take a 4-dimensional, non-degenerate `initial'  metric $g_{\textrm{\textsc{in}}}(x^\mu)$ on spacetime $M$. This step essentially involves the Penrose conformal method \cite{pen86}.
\item Conformally deform $g_{\textrm{\textsc{in}}}$ to a new metric $g_4=\Omega^2g_{\textrm{\textsc{in}}}$  by choosing a suitable conformal factor $\Omega$. This step connects the `bad' metric $g_{\textrm{\textsc{in}}}$ with the `nice', non-degenerate, and non-singular  metric $g_4(x^\mu)$.
\item Using the method of asymptotic splittings for the 5-dimensional Einstein equations with an arbitrary (with respect to the fluid parameter $\gamma$) fluid (\ref{ein}),  solve for the 5-dimensional metric $g_5=a^2(y)g_4+dy^2$ and the matter density  $\rho_5$.
\item Transform the solutions of step 3 to suitable factored forms of the general type, (divegent part) $\times$ (smooth part).
\item  Construct the `ambient' metric in normal form, $g_+$, for the 5-dim Einstein equations (\ref{ein}) with a suitable fluid.
\item $(M,[g_4])$ is the conformal infinity of $(V,g_+)$, that is   $\mathscr{I}=\partial V=M$.
\item The metric $g_+$ is conformally compact. This means that a suitable metric $\mathring{g}$ constructed from $g_+$ extends smoothly to $V$, and its restriction to $M$, $\mathring{g}|_M$, is non-degenerate (i.e., maintains the same signature also on $M$, cf. \cite{fef} for this definition).
\item The conformal infinity $M$ of the ambient metric $g_+$ of any metric in the conformal class $[g_4]$ is controlled by the behaviour of a constant rescaling of the `nice' metric $g_4$.
\item As a conformal manifold, $(M,[g_4])$ has no singularities. This means that there is always a regular metric on $M$: the metric $\mathring{g}|M$  belonging to $[g_4]$ is regular. (See Section 7.)
\item Cosmic censorship on $(M,[g_4])$ is equivalent to the validity of the asymptotic condition satisfied by  the ambient metric $\mathring{g}|_M$. (See Section 8.)
\end{enumerate}
Some explanatory remarks  are in order here. The initial 4-metric $g_{\textrm{\textsc{in}}}$ of step 1 is any metric with a conformal infinity. For example, de Sitter, Anti-de Sitter, or Minkowski metrics, when written is suitable forms using the conformal method \cite{pen86}, show explicitly their conformal infinities. Similarly, for the Schwarszchild, Reissner-Nordstrom, or Kerr metrics,  $g_{\textrm{\textsc{in}}}$ represents their maximal analytic extensions (e.g., the Kruskal, double-null forms, etc). The metric $g_{\textrm{\textsc{in}}}$ written in this way, has a regular region which (like in the Einstein static universe for the exact models) is bounded by  certain hypersurfaces,  the conformal boundary. This boundary represents the various infinities of the metric, but most importantly also contains the singularities (in the sense of the classical singularity theorems \cite{he}), previously located at some finite point not at infinity.

For example, in the FRW spaces, the bounding hypersurfaces of the metric $g_{\textrm{\textsc{in}}}$ represent points at infinity,  but also contain the various  singular points as some finite part of the conformal boundary at infinity. So for instance the singularity at $t=0$ is  represented by a spacelike surface (generally part of the bounding hypersurfaces) at infinity. This is the meaning of the conformal method \cite{pen86}, and it applies similarly to other exact spacetimes  (see also \cite{he}). This is very convenient for what follows because a $t=0$, proper time singularity surface, which would prevent us from comparing the situation with a better metric in connection with the ambient construction, has now moved to infinity, and we can proceed with our 5-dimensional constructions even for singular 4-metrics (see below and especially Section 7).

In step 2, the term  `nice' has a suitable technical meaning  giving better asymptotic properties over the metric $g_{\textrm{\textsc{in}}}$ that we started with. In this sense,  considering the conformal form of  spacetimes in step 1 and having the classical singularities at the conformal boundary by that construction, has the advantage that we may basically kill off the infinities and singularities of the $g_{\textrm{\textsc{in}}}$  metric simply by multiplying by a suitable conformal factor, and so end up with a  regular spacetime, $g_4$, for instance, Minkowski or de Sitter or Anti-de Sitter. In all cases, we will have a boundary $M$ possessing a conformal structure $[g_4]$, where any two metrics belonging to  $[g_4]$ are conformal transformations of each other.  For instance,  taking $g_{\textrm{\textsc{in}}}$ to be the flat FRW metric in its conformal form, or an exact  black hole spacetime (like the  maximally extended spacetimes mentioned above),  we may choose a suitable, smooth conformal factor in step 2, and obtain the Minkowski metric for $g_4$.


We also note that in steps 1-2, we do \emph{not} impose any dynamical equations for the metric $g_{\textrm{\textsc{in}}}(x^\mu)$ or $g_4$. In this way, we avoid the usual problems of having to decide which of the two, conformally related metrics represents a `physical' metric among the two conformal `frames' (cf. \cite{dicke} and references therein). It is an important feature of the ambient construction that in the end, for any  two conformally related, 4-dimensional  metrics belonging to the conformal structure $[g_4]$ of $M$, their ambient metrics restricted on the boundary $M$ can be chosen to be not much different than the restricted ambient metric corresponding  to $g_4$.

In step 3, the $g_4$ metric is the one from step 2. It may also be $g_{\textrm{\textsc{in}}}(x^\mu)$ from step 1, provided we exclude the points with coordinates $x^\mu$ on the singular boundary where the metric becomes infinite. One may wonder whether we are allowed to exclude such `ideal' points from the metric $g_{\textrm{\textsc{in}}}(x^\mu)$ in the sought-for 5-dimensional ansatz $a^2(y)g_{\textrm{\textsc{in}}}(x^\mu)+dy^2$, because such 5-dimensional solutions are obviously not defined there. However, since we do not impose any field equations on $M$, we do not know the domain of definition of the metric $g_{\textrm{\textsc{in}}}(x^\mu)$, even locally,  starting from initial data on $M$. Therefore the only way to proceed is if we `move' all possible singular finite points of the metric $g_{\textrm{\textsc{in}}}(x^\mu)$ to become ideal points at infinity, so that $g_{\textrm{\textsc{in}}}(x^\mu)$ is regular elsewhere on $M$. This is the purpose of using the conformal method on $g_{\textrm{\textsc{in}}}(x^\mu)$.
Then, in step 3, we build asymptotic 5-dimensional solutions in the form of formal series expansions using the technique of asymptotic splittings \cite{skot}. This method produces solutions for the warp factor and the density of the form $a(y)=y^{p_1}\sum_{i=0}^{\infty}c_iy^{i/s},\; \rho(y)=y^{p_2}\sum_{i=0}^{\infty}d_iy^{i/s},\;y\rightarrow 0.$  All such solutions are singular in a finite distance $y_s$ from the boundary  located at $y=0$ in the variable $y-y_s$. For notational simplicity we  set $y_s=0$ here and in everything below, but this has the apparent effect of moving the singularity from the point at finite distance $y_s$ from the boundary,  where it was originally located, to the boundary of the ambient space itself at $y=0$. We also note that according to the  method of asymptotic splittings, the denominator $s$ appearing in such expansions is the same for \emph{all} components of the  solution $x(y)=(a,\dot{a},\rho)$ of the  dynamical system  $\dot{x}=f(x)$ representing the 5-dimensional Einstein equations with the aforementioned fluid source.

Steps 4, 5 are studied  in Section 4. In step 4, we show how to convert  solutions of step 3 to the factored forms, $a(r)= r^{\kappa_1}\,\xi(r),\rho(r)=r^{\kappa_2}\,\zeta(r)\; \textrm{as}\; r\rightarrow 0$, which contain a divergent, power-law  part, and a smooth, \emph{convergent} part. It is curious that this step necessarily introduces an asymmetry in the $y$-dimension, in the sense that we no longer can have invariance under the symmetry $y\rightarrow-y$ in the extra dimension $y$. The transformation that prepares the solutions to be brought to a suitable form that implies the convergence we need (so that we may apply various relevant theorems, see the first half of Section 4), is  a very simple change in the parametrization of the extra dimension, $r=r(y)$, which itself has certain properties: 1) It is consistent with the (later) property that the 5-dimensional spacetime has a conformal boundary  at $y=0$, 2)  it is well-defined only in the half-space $y>0$\footnote{This is similar to the situation with Minkowski space being the boundary of the AdS space in the AdS/CFT correspondence.}, 3) the boundary in the new parameter $r$ is again located at 0, and 4)  it strongly highlights the particular  features of the series solutions we use.

We continue, in the second half of Section 4, with  step 5: The ambient metric has the form $g_+=w^{-n}\left(\sigma^2(w)g_4(x^\mu)+dw^2\right), n\in\mathbb{Q}^+$, as $w\rightarrow 0$, with $\sigma(w)$ a smooth (infinitely differentiable) function such that $\sigma(0)$ is a nonzero constant. The metric $g_+$ (essentially $g_5$ up to constant rescalings in different variables) solves the 5-dim Einstein equations (\ref{ein}) with a suitable fluid with density $\rho_+(w)$ that has the generic form $\rho_+=w^{-z}\theta(w)$, $z\in\mathbb{Z}$ and $\theta(w)$ smooth. Both functions $w(r)$ and $w(y)$ are well-defined and invertible.

Step 6 is studied  in Section 5 after we have introduced the metric $\mathring{g}$. For this step, we prove that when $\mathring{g}$ is restricted on $M$, it becomes a metric in the conformal class $[g_4]$. In the first part of Section 5, we also complete the construction of the ambient metric started in  the previous section, for two special cases which need for their treatment the introduction of the metric $\mathring{g}$.

The property of conformal compactness of step 7 is shown later in Section 5. It means that  we can always pass to the metric $\mathring{g}=\omega(w) g_+$, $\omega$ smooth (in a slight abuse of language, we will also call $\mathring{g}$ the ambient metric), that extends smoothly on $V=M\times\mathbb{R}$, and when restricted to the first factor $M$  gives the metric $\mathring{g}|M$ that is non-degenerate. Although $\mathring{g}$ does not solve the Einstein equations like $g_+$, they both share the same conformal boundary and so everything we say about the behaviour of $\mathring{g}$ on $M$ holds essentially also for $g_+$.

For step 8 we show in Section 6 that for a metric $g\in[g_4]$, the difference of the two ambient metrics $\mathring{g}-\mathring{g}_4$ has a restriction on $M$ which satisfies a set of asymptotic conditions. These conditions imply that the ambient metric  $\mathring{g}|_M(0)$, for any $g\in[g_4]$, is a constant rescaling of $g_4$, and therefore has an improved behaviour over the original 4-metric $g$.

We thus end up with the following situation. There is  a ambient manifold $(V,g_+)$ on which we have the Einstein equations (\ref{ein}) valid for the ambient metric $g_+$ as defined above. The ambient spacetime has a boundary where only a conformal structure $[g_4]$, not a unique metric, is defined. The construction is local in nature and shows that the asymptotic structure of the ambient 5-dimensional geometry $g_+$ at its conformal boundary  $(M,[g_4])$ has  conformal infinity the metric $\mathring{g}|_M$ as defined above. For any given 4-metric $g\in[g_4]$, its ambient 5-metric is given by
\be
g_+=w^{-n}(g_w+dw^2),
\ee
where
\be
g_w= \sum_{i=0}^{\infty}c_iw^i g,\quad w\rightarrow 0,
\ee
with the coefficient of $g$  being a convergent formal power series of $w$, such that $g_0=c_0g$ ($c_0=\sigma(0)$ is  a nonzero constant). Hence,
 $g_w$  represents a 1-parameter family of boundary metrics constructed recursively by the method of asymptotic splittings, and is asymptotic to $\sigma^2(w)g$, where $\sigma^2$ is the asymptotic smooth sum of the power series. This sum, in the limit $w\rightarrow 0$, gives the metric $\mathring{g}|_M(0)$ belonging in the conformal structure of the boundary $[g_4]$ and having constant conformal factor.

This construction has implications for singularities and the question of cosmic censorship on the boundary 4-spacetime $M$ as we discuss in Sections 7, 8.

\section{The normal form of the ambient cosmology}
It is not immediately obvious that a general Fuchsian  formal  series  leads to a well defined function; in fact, such a series generally will not because it either contains \emph{general} powers of the form $\sum_{i=-\infty}^{\infty}c_iy^{\eta_i},\eta_i\in\mathbb{Q}$, or log terms, or both, cf. \cite{go}.

However, in this section we  show that due to its special form, that is having \emph{equal }denominators (all being equal to $s$), the series factor in Eq. ({\ref{scale})  converges provided we introduce an asymmetry in the $y$ dimension, and this gives an important first step to the realization of our basic construction in this and the following sections.

To begin, we introduce the parameter
\be\label{r}
r=y^{1/sl},\; p=m/l,\,m\in\mathbb{Z},l\in\mathbb{N},
\ee
into (\ref{scale}) and notice that  this is real provided we choose $y>0$,  and well-defined since $s,l$ are positive integers by definition. This  implies that we can have no symmetry of the form $y\rightarrow-y$ in $V$. Then it follows that (\ref{r}) is an invertible transformation because $dr/dy=(1/sl)y^{1/sl-1}>0$.

Since $r\rightarrow 0$ when $y\rightarrow 0$,  the Puiseux series (\ref{scale}) then trivially becomes a formal power series in disguise,
\be\label{scale2}
a(r)=r^{sm}\sum_{i=0}^{\infty}c_ir^{il},\quad r\rightarrow 0.
\ee
In this new form the \emph{convergence} of the series in (\ref{scale2}) follows because according to a theorem of Borel (cf. Ref. \cite{rem}, p. 300), when the coefficients $c_i$ are real, the formal power series $\sum_{i=0}^{\infty}c_ir^i$ always converges to a smooth function $\sigma:I=(-\epsilon,\epsilon)\rightarrow \mathbb{R}, $ for some $\epsilon >0$. A basic point in the proof of Borel's theorem is to use the mean value theorem of differential calculus and the constants $c_i$ to extend the derivative  $\sigma^{(n)}$ as a continuous function \emph{everywhere} in $I$ by assigning it, at $0$, the value $\sigma^{(n)}(0)=c_n,n=0,1,2,\cdots$ The resulting function $\sigma$ will also be real analytic in a small neighborhood around each nonzero point of the  interval $I$.
Therefore the warp factor will assume the form\footnote{Actually, the theorem of Borel  is a special case of Ritt's theorem about the convergence of formal power series with \emph{complex} coefficients, cf \cite{rem}, p. 299. Such a power series will always be asymptotic to an analytic function $\sigma$ on an angular-shaped  sector $\mathcal{S}$,  that is we have $\sigma (z)\sim\sum_{i=0}^{\infty}c_iz^i,\,z\rightarrow 0$, on $\mathcal{S}.$ }
\be\label{scale3}
a(r)= r^{\kappa}\,\sigma(r),\quad \textrm{as}\quad r\rightarrow 0,\quad \kappa=sm\in\mathbb{Z},
\ee
so that $a$ will always be the product of some smooth (in fact real analytic) function times the factor $r^{\kappa}$. We note that because of  the $y$-asymmetry we must necessarily have that $r>0$.

Now since $p$ is a parameter in the theory (determined by the structure of the field equations and the method of asymptotic splittings), there is the issue of understanding the physical significance of the solutions as $p$ varies and, in particular, in the limit when $p\rightarrow\infty$ in (\ref{scale}).  We showed above that when $p$ is bounded, the warp factor reduces to the canonical form (\ref{scale3}), and so the 5-dimensional geometry (\ref{metric1}) becomes
\be\label{q}
g_5(r)=r^{2sm}\left(\sigma^2g_4+r^{-2q}dr^2\right),\quad q=sm-sl+1.
\ee
We note here that $q\neq 1$, iff $m\neq l$ (i.e., $p\neq 1$). In this form, $\sigma$ is a new formal power series with coefficients determined recursively by the method of asymptotic splittings. We use  again the same letter $\sigma (r)$, but now $r$ stands for the rescaled variable $r/sl$ which is also necessarily positive, and there is an overall additional constant factor  $(1/sl)^{2sm}$ multiplying the brackets (again not shown for simplicity). To get $\sigma$, one considers
the dominant part $f^{(0)}$ of the vector field, and the
 eigenvalues of the Kowalevskaya matrix $K=Df^{(0)}(A)-\textrm{diag} P$ (where the dominant balance is $At^P$) which are rational numbers, and  where $s$ is defined to be the least common multiple of their denominators.

When  the exponent $q$ defined in Eq. (\ref{q})  is  different from one we may introduce  a new parameter $w$ (not necessarily positive this time), with \be dw=r^{-q}dr.\ee Because $y>0$, we have that $dw/dr=y^{-q/sl}$ and $dw/dy=(1/sl)y^{-p}$ are  positive, and so both reparametrizations are invertible.
Then the 5-dimensional geometry (\ref{q}) becomes
\be\label{metric2}
g_+=w^{-n}\left(\sigma^2(w)g_4(x^\mu)+dw^2\right),\quad w\rightarrow 0,
\ee
where
\be\label{n}
-n=\frac{2m}{l-m}=\frac{2p}{1-p}.
\ee

We call $g_+$, which stands for $g_5(w)$ (possibly rescaled by a constant factor of the form $(1-q)^{-2sm/(1-q)}$ if necessary), the normal form of the \emph{ambient} metric. The exponent $n$ satisfies the following properties:
\begin{itemize}
  \item $1-p\neq 0$
  \item $-n<0$
  \item $n=n(p)$, and in fact, $-n\neq -2$
  \item $-n\rightarrow -2$,  as $p\rightarrow\infty$.
\end{itemize}
Before we discuss the properties of this enumeration, we note here that there are two cases, for $p=1$ (already excluded from the discussion above when we assumed that $q\neq 1$) and when  $p\in(0,1)$ (when $w\rightarrow 0$), that are not included in the above discussion. For these two cases,  the prefactor in the normal form of the ambient metric $g_+$ in Eq.(\ref{metric2}) tends to zero and is not divergent. We shall discuss these two cases separately in the next section, when we introduce the metric $\mathring{g}$.

Returning to the enumeration above, the first property means that the exponent in the ambient metric (\ref{metric2}) is always well-defined. We  note that $\sigma$ is  a new smooth (in fact, real analytic) function, for which we keep the same symbol as before.
The second property justifies the name `normal form' which we gave to the ambient metric (\ref{metric2}), in analogy to the Fefferman-Graham (FM) ambient metric construction \cite{fef}. Their brilliant work provides an existence theorem for the ambient metric in the case of the  5-dimensional Einstein equations with a cosmological constant, and they also show that their ambient metric $g_+=w^{-2}(\sigma^2g_{4}+dw^2)$ is essentially unique.
However, the third property of $n$ implies that our ambient metric is distinct from the FM metrics (the latter always have the constant  characteristic exponent equal to $-2$ in place of our $-n(p)$ in (\ref{metric2})). This has a  novel consequence for our metrics $\mathring{g}$ (to be introduced shortly below), that instead of having uniqueness for $g_+$, we necessarily find a nontrivial asymptotic condition on the scri of the ambient metric $g_+$ for $\mathring{g}|_M$. The last property of $n$ in the enumeration above implies that the FM metrics appear in this context as the $p\rightarrow +\infty$ limit of our geometry.

The form of the ambient metric (\ref{metric2}) implies that the original (asymptotic splitting) form of the density  $\rho =y^{p'}\sum_{i=0}^{\infty}c_iy^{i/s}$, with $p'=m'/l'\in\mathbb{Q},m'\in\mathbb{Z}, l'\in\mathbb{N}$, becomes  in terms of $r$,
 \be
 \rho(r)=r^{sm'}\sum_{i=0}^{\infty}c'_ir^{il'},
  \ee
 and therefore it takes the generic form (for $p\neq 1$)
\be
\rho_+=w^z\theta(w),\quad z=\frac{p'}{1-p}\frac{l'}{l}, \quad\theta\; \textrm{smooth}.
\ee
When $p=1$, we find, $\rho (w)=e^{m'w}\theta(w)$, and then the transformation $w\rightarrow 1/w$ makes the density a $C^{\infty}$ function,  if we define $\rho(0)=0$.

\section{The conformal infinity of the ambient metric}
To complete the construction of the ambient metric of the previous  section, we  discuss the two cases we left previously, namely, when $p=1$  and when  $p\in(0,1)$. To proceed, on $V=M\times\mathbb{R}$, we define the product metric
\be\label{best metric0}
\mathring{g}=\omega^2 g_+,
\ee
where $\omega>0$ on $V\setminus M $, and zero on $M$. (Later in this section,  we choose $\omega=w^{n}$, but this is not important presently.)
We will shortly see that in this form,  the metric $\mathring{g}$ in Eq. (\ref{best metric0})  represents the regular (see below) conformal infinity of $g_+$, and justifies step 6 of Section 3. This means that when $g_+$ blows up to $+\infty$ at points of the boundary $M$, the conformal factor $\omega$ tends to zero and so is responsible for a conformal squashing of $g_+$ producing a finite $\mathring{g}$ at infinity, whereas when $g_+$ crunches to zero, $\omega$ blows up to infinity, again making  $\mathring{g}$ finite there. All results above correspond to $\omega\rightarrow 0$ at the conformal infinity of $g_+$, since $g_+$ diverged there, but when $p=1$, or $p\in(0,1)$, we shall find that $\omega\rightarrow\infty$.

When $p=1$, the ambient metric will read $g_+=e^{2w}(\sigma^2(w)g_4(x^\mu)+dw^2)$, where $w/s=\ln r$, and so at the conformal boundary (i.e., as $r\rightarrow 0$), the conformal factor of the ambient 5-geometry in Eq. (\ref{best metric0}), $\omega^2=e^{-2w}$, will tend to $+\infty$ at $\mathscr{I}$. Similarly, when $p\in(0,1)$, then $-n>0$, and the ring metric becomes $\mathring{g}=w^ng_+$, which means that the conformal factor $\omega =w^n,n<0$,  diverges at the boundary ($w\rightarrow 0$ now). We therefore see that both these cases are exceptional in the sense that the conformal factor $\omega$ diverges at the boundary points, instead of tending to zero there as it does for all other $p$ values. Hence, it appears that when $p\in(0,1]$, the behaviour of $\omega$ which describes the scaling between the two metrics $\mathring{g}$ and $g_+$ is somehow inverted, so that $\omega$ becomes its reciprocal, $\omega\rightarrow1/\omega$.

Since the exponent $p$ is generically \cite{ack} a function of the equation-of-state parameter $\gamma$ in the 5-fluid equation of state $P=\gamma\rho$ in Eqns. (\ref{scale})-(\ref{ein}), we may interpret the above situation by adopting the view that $\omega$ (or its inverse) as a function of $\gamma$ is not smooth. In general,  $\omega$ is also  a real-valued function defined on the set $M\times\mathbb{R}$, so we may write $\omega=\omega_\gamma(x^\mu,w)$, and think of the conformal factor $\omega$ as a 1-parameter family of functions on $V$. The description of the conformal boundary $M$ in terms of properties of the family $\omega_\gamma$ is a more complicated problem, currently under investigation.

Having constructed the ambient metric $g_+$ given by (\ref{metric2}),  steps 6, 7 of the construction  now  follow. For, if on $V=M\times\mathbb{R}$ we define the product metric
\be\label{best metric}
\mathring{g}=w^{n}g_+=\sigma^2 g_4+dw^2,
\ee
and since $g_4$ is non-singular, we find that
\be
\det{\mathring{g}}\neq 0, \quad \sigma^2(w)=c_0+c_1w+\cdots,
\ee
so the metric (\ref{best metric}) is manifestly regular at $w=0$. Also, the metric $(w^{n}g_+)|_M=\sigma^2(w) g_4$ is clearly a metric belonging in the conformal class $[g_4]$ (albeit one with smooth conformal factor not depending on $x^\mu$ but only on  $w$). Thus a conformal infinity exists for the ambient metric $g_+$ given by (\ref{metric2}), and step 6 follows.

For the conformal compactness of the ambient metric $g_+$ of step 7 of Section 3, we first note that since  $\sigma(w)$ is a smooth function, we find that (\ref{best metric})  extends smoothly everywhere in the ambient spacetime $V$. As we already discussed, the metric $\mathring{g}$ is non-unique in the sense that any  function $\omega(w)$ with the required properties will do. Here, we have chosen $\omega=w^n$, but this choice is not unique.

To show the non-degeneracy part of the property of conformal compactness, we proceed as follows. Since $\mathring{g}$ is a spacetime metric, it will have indefinite signature. Because its restriction to the first factor, $\mathring{g}|_M$, that is the pullback  $j^*(\mathring{g})$ by the inclusion map $j:M\rightarrow V$, is given by the form
\be\label{pull}
\mathring{g}|_M= \sigma^2(w) g_4,
\ee
it follows that $\mathring{g}|_M$ will be a well-defined metric on $M$ if and only if $g_4$ is non-degenerate, so that $\mathring{g}$ sustains  the same signature also on $M$. Now, the Penrose construction \cite{pen86} utilized in step 2 generally guarantees the nondegeneracy of the `nice'  metric $g_4$. In general, the non-degeneracy of the metric $\mathring{g}|_M$ dictates that $g_4$ should not become degenerate at any point of $M$, in any case,  we view (\ref{pull}) as a kind of compatibility condition.  We shall have to say more on this in Section 8.

To prepare for step 8 and the asymptotic conditions treated  next, we end this section with the following remarks.
From the results up to now, it follows that for two conformally related 4-metrics of $M$, $g_1=\Omega^2g_2$, the difference of their ambient metrics on $M$ is at best (meaning even for $w=0$) equal to  $\mathring{g}_1|_M(0)-\mathring{g}_2|_M(0)=c_1g_1-c_2g_2,$ with $c_1,c_2$ being the well-defined, nonzero constants $c_i=\sigma^2_i(0), i=1,2$. So it is not obviously zero, and uniqueness of the ambient metric cannot really follow by simply quoting the case of the FG metrics (cf. \cite{fef}, chapters 3, 4). This is because in their metrics in normal form, $g_+=w^{-2}(g_w+dw^2)$, the 1-parameter family $g_w$ of metrics on $M$ does not necessarily have the same form as our explicitly constructed $\sigma^2(w) g$. In any case, they do not have a fluid in the ambient space like we do, only a cosmological constant, and so an  adaptation of their uniqueness results to our case is not straightforward. We also note that while the property of being an Einstein space, i.e., $\textrm{Ric}(g_4)+\lambda g_4=0$, is preserved under a conformal rescaling with constant factor, $g_4\rightarrow cg_4$, this is not so for solutions of the Einstein equations with a fluid source.

\section{The asymptotic conditions}
In this section we focus on proving step 8 of Section 3, in particular,  we show that the ambient metric of any 4-metric in the conformal  class of a given 4-metric on $M$,  although not unique, has a `universal'  conformal infinity  in the sense of giving a hugely simplified geometry in comparison to the initial 4-metric on $M$  it arose from.

In particular, we show that if  we take two 4-metrics $g_1,g_2$ on $M$ belonging to the same conformal class, that is  such that there is a smooth conformal factor $\Omega(x^\mu)$ with $g_1=\Omega^2g_2$, then the $M$-restriction of the ambient metric of $g_2$ at $w=0$, that is the metric $\mathring{g}_{2}|_M(0)$, equals a nonzero constant times the `good' metric $g_1(x^\mu)$. We may think of $g_1$ as a `good' metric as in step 2 and $g_2$ as a `bad' metric of step 1 - i.e.,  one blowing up at $\Omega=0$, the scri of $M$,   $\mathscr{I}_M$.

For two such 4-metrics $g_1,g_2$ of $M$, their ambient metrics on $M\times\mathbb{R}$ are given as the product metrics
\be\label{two}\mathring{g}_i(w)=\sigma^2_i(w)g_i(x^\mu)+dw^2,i=1,2. 
\ee
Since $g_2(x^\mu)$  would possibly become infinite at points of its conformal boundary, we exclude such points $x^\mu$ from the right hand side of the equation (\ref{two}) for $\mathring{g}_2(w)$. Then
the restrictions to $M$ of the two ambient metrics in (\ref{two}) will differ by the symmetric 2-tensor   $\hat{g}(w)=\mathring{g}_1|_M(w)-\mathring{g}_2|_M(w)$ where,
\be\label{w}
\hat{g}(w)=g_2\left(\sigma^2_1(w)\Omega^2(x^\mu)-\sigma^2_2(w)\right),
\ee
and the problem becomes one of deciding how bad this difference can really be. This difference measures how far the ambient metric of the `bad' metric $g_2$ is from that of the good metric $g_1$. If this difference were not near $g_1$, then the ambient construction could not really be considered as something worth doing. On the other hand, if this difference were zero then the ambient metric would be the one and only metric for all 4-metrics in the conformal class of a given 4-metric on $M$. This is the situation with the FG metrics \cite{fef}. Below, we show that in our construction this difference, although nonzero, becomes a constant multiple of $g_1$, and so the ambient metric of any  $g_2$ in the conformal class of $g_1$ corresponds to a great simplification over the original situation of having the two metrics $g_1,g_2$ on $M$.
We note that the difference $\hat{g}(w)$ is not a well-defined metric on $V$ because it is degenerate, and so its projection to $M$ is  not defined. Only when first restricting the two ambient metrics (\ref{two}) to $M$ as in (\ref{pull}) and then taking their difference, produces a non-degenerate, well-defined tensor field   on $M$ (see also below).

Setting $w=0$ in Eq. (\ref{w}),  the difference will be,
 \be\label{con}
 \hat{g}|_M=g_2(c_0\Omega^2(x^\mu)-c_0'),
  \ee
  where $c_0,c_0'$ are the first constants in the formal series of the two ambient metrics. Since these constants are nonzero and uniquely determined by the method of asymptotic splittings, while $\nabla_{g_1}\Omega\neq 0$, as this is assumed in the conformal method \cite{pen86}  so that $\Omega$ is not a constant on $M$, the term in the brackets in Eq. (\ref{con}) is not zero. Therefore, the function in the brackets in Eq. (\ref{w}) is not identically zero. Hence, we may  regard this  $\hat{g}(w)$ from Eq. (\ref{w}) as a well-defined, 2-tensor living on the original 4-manifold $M$, that has the variable $w$  as a \emph{parameter}. In this  sense,  $\hat{g}(w)$ represents a 1-parameter family of 4-dimensional tensor fields on $M$ depending on $w$, and so we can take  successive derivatives of $\hat{g}(w)$ in Eq. (\ref{w}) with respect to $w$ and then set $w=0$.

Now, we know from the method of asymptotic splittings that at  finite orders in the series $\sigma_1,\sigma_2$, there appear \emph{arbitrary} constants as coefficients precisely in those terms in the series where the exponents equal the eigenvalues of the Kowalevskaya matrix. So if the order $i=i_0$ is such that $i_0/s=\rho_{i_0}$,  where $\rho_{i_0}$ is an eigenvalue of the  Kowalevskaya matrix (for the 5-dimensional Einstein equations with the fluid source we solved to find $g_+$), then the corresponding coefficient $c_{i_0}$ in front of that term in the series $\sigma(w)=\sum_{i}^{\infty}c_iw^i$ will be an \emph{arbitrary} constant. We know that arbitrary constants appear in the terms with exponents equal to those eigenvalues of the Kowalevskaya matrix (the so-called `K-exponents') which have strictly positive real parts (corresponding to the unstable eigenspaces), and this property is not only restricted to the case of the Puiseux series (i.e.,e rational exponents), but continues to hold for the general $\Psi$-series containing log terms. Provided  we set the arbitrary  coefficients corresponding to the negative eigenvalues equal to zero, the resulting series developments always exist as convergent sums \cite{go}.

We note in passing that our restriction to Puiseux series from the beginning in this work was because we looked  for \emph{general} solutions to the 5-dimensional Einstein equations with fluid sources in \cite{ack} in an effort to answer the question whether or not \emph{generic} solutions of the system were singular, that is whether singularities were a typical feature of all solutions in 5-dimensions. Had we instead started with a $\Psi$-series ansatz containing log terms and set to zero some of the arbitrary constants in the supposed  solutions at the end would produce only particular solutions, since we would have sacrificed some of the arbitrary constants, and could not contribute usefully to\emph{ that }search.  However, for the purposes of the arguments in the current work, we only need \emph{some} (in particular, the positive) arbitrary constants to be nonzero, not all, and therefore we conclude that all arguments presently would continue to be valid had we considered a $\Psi$-series instead of simple Puiseux expansions from the beginning.

We can now  differentiate Eq. (\ref{w}) with respect to $w$ precisely $i_1$ times, where $i_1$ is the order where the first arbitrary coefficient $c'_{i_1}$  in the expansion $\sigma^2_2(w)$ associated with the \emph{`bad'} metric $g_2$ appears. Then Eq. (\ref{w}) becomes
\be
\frac{\partial^{\,i_1} \hat{g}}{\partial w^{{i_1}}}(w)=g_2\left((c_{i_1}+c_{i_1+1}w^{i_1+1}+\dots)\Omega^2(x^\mu)-(c_{i_1}'+c'_{i_1+1}w^{i_1+1}+\dots)\right).
\ee
Here $c_{i_1}$ is fixed but the constant $c_{i_1}'$ is arbitrary. This is so because the Kowalevskaya matrices corresponding to the metrics $g_1,g_2$ do not in general have the same eigenvalues, so that one should not expect that \emph{both} constant coefficients  $c_{i_1},c_{i_1}'$ will be arbitrary at the same order $i_1$. Setting now $w=0$, we find that
\be\label{ac-}
\frac{\partial^{\,i_1} \hat{g}}{\partial w^{{i_1}}}\Arrowvert_{w=0}=g_2\left(c_{i_1}\Omega^2(x^\mu)-c_{i_1}'\right).
\ee
In this equation, unlike Eq. (\ref{con}), we are allowed to set the arbitrary constant $c_{i_1}'=0$. Then we find the following asymptotic condition,
\be\label{ac}
\frac{\partial^{\,i_1} \hat{g}}{\partial w^{{i_1}}}\Arrowvert_{w=0}=c_{i_1}g_1(x^\mu),\quad \textrm{on}\quad M,
\ee
where   $g_1$ is the 4-dimensional  metric on $M$ with `nice' properties. We note the following remarkable fact: Although from the beginning of this Section and up to and including Eq. (\ref{ac-}) we had to exclude the singular boundary points where  the `bad' metric $g_2$ would diverge,  Eq. (\ref{ac}) holds \emph{even} at those singular points because $g_1$ is perfectly regular there.

Now suppose that the Kowalevskaya matrix corresponding to the 5-dimensional Einstein equations for ambient metric $g_{2,+}$ of the \emph{`bad'} metric $g_2$ has exactly $k$ positive eigenvalues $\rho_{i_1}=i_1/s,\cdots\rho_{i_k}=i_k/s$. Following the procedure above, we will end up with $k$ asymptotic conditions, one  for each one of the  $k$  derivatives of various orders of the metric $\hat{g}$, evaluated at $w=0$:
\be\label{set}
\frac{\partial^{\,i_1} \hat{g}}{\partial w^{{i_1}}}\Arrowvert_{w=0}=c_{i_1}g_1,\cdots, \frac{\partial^{\,i_k} \hat{g}}{\partial w^{{i_k}}}\Arrowvert_{w=0}=c_{i_k}g_1,\quad \textrm{on}\quad M.
\ee
This is the set of the required asymptotic conditions we need to determine $\hat{g}$. For, since we know that $k$ derivatives of that function at $0$ are all constants times the metric $g_1$, we conclude that
\be\label{w1}
\hat{g}|_M(w)=e^{aw}g_1(x^\mu),\quad a=\sum a^jc_j,\; j={i_1},\cdots,{i_k}.
\ee
The $k$ unknown constant coefficients $a^j$ are  found by solving the nonlinear algebraic system 
\be\label{alg}
\left(\sum a^jc_j\right)^{i_1}=c_{i_1},\cdots,\left(\sum a^jc_j\right)^{i_k}=c_{i_k}.
\ee 
This system always has a solution for $a$ in the complex field. We note that Eq. (\ref{w1}) holds also for the singular boundary of the `bad' metric $g_2$ for the same reason as before, namely, because $g_1$ is perfectly regular at these points.

We may think of the result (\ref{w1}) as valid in a small, local neighborhood around $w=0$, because the extension of the conformal boundary of the 4-metric $g_1$ to $V$ is regular since we proved that $\mathring{g}_1$ is regular in $V$. There, the difference $\hat{g}|_M(w)$ is a well-behaved multiple of $g_1$, the factor depending only on $w$. Setting $w=0$ in Eq. (\ref{w1}), we find that
\be\label{ac*}
\mathring{g}_1|_M(0)-\mathring{g}_2|_M(0)=g_1,
\ee
essentially a condition on the conformal infinity of the ambient metric $g_{2,+}$ of the `bad' 4-dimensional metric $g_2$. It
means that the conformal infinity of the 5-metric $g_{2,+}$
is  equipped with the metric $\mathring{g}_{2}|_M$, which  differs from the `nice' conformal infinity $\mathring{g}_{1}|_M$ by  the `nice' metric $g_1$,  and so it cannot be very `wild', it will be a  rescaling of the `good' metric $g_1$. This is true also  for the points on the singular 4-boundary of the metric $g_2$ for reasons discussed above.

This $w$-rescaling of $g_1$ from Eq. (\ref{w1}) will have the form $\sigma^2_1(w)-e^{aw}$, where $\sigma^2_1$ is a convergent, formal power series expansion for the ambient metric $g_{1,+}$ of the `good' metric $g_1$, with $w$ taking values in a small neighborhood of the form $(-\epsilon,\epsilon)$ around 0. At 0, it will read $c_0-1$, and so we find
\be\label{set1}
\mathring{g}_{2}|_M(0)=(c_0-1)g_1(x^\mu).
\ee
We note that when $c_0$ turns out to be  1 (as it could come out by the application of  the method of asymptotic splittings), we may use the freedom to multiply the right hand side of Eq. (\ref{w1}) by  a different integration constant of the form $C=e^b$, which was chosen equal to 1 above. The system for $a=\sum a^jc_j+b$ would then solve as before, but with $b$ added inside the brackets in the left hand sides of (\ref{alg}), namely, we will have the system,
\be
\left(\sum a^jc_j+b\right)^{i_1}=c_{i_1},\cdots,\left(\sum a^jc_j+b\right)^{i_k}=c_{i_k}. 
\ee
Hence the constant in the right hand side of (\ref{set1}) may be assumed to be nonzero.
 When this factor  is positive, the constant rescaling (homothety) of the metric $g_1$  preserves the causal character of curves on $M$, while when it is negative we have the so-called \emph{metric-reversing} \cite{o'neill}. In this case, all geometric notions associated with the homothetic transformation remain the same, however, the causal character of timelike and spacelike vector fields is reversed, while null ones remain so. (In any case, the sign of the constant factor is controlled by the choice of $b$.)

 Eq. (\ref{set1}) is then, in our context, the condition replacing uniqueness of the Fefferman-Graham ambient metric. We call  this `the asymptotic condition' because it is one valid on the conformal infinity $M$ of the ambient space $V$ after taking the limit $w\rightarrow 0$. We have shown the following result.
\begin{theorem}
A  4-metric $g$ in the conformal class of a `good' 4-metric $g_4$ on $M$ has an ambient metric $g_+$ which satisfies the 5-dimensional Einstein equations with a fluid source and has conformal infinity $(\mathscr{I}_{g_{+}},\mathring{g}|_M)$  described as  a  constant rescaling of $g_4$, $\mathring{g}|_M=cg_4$. Any two conformally related 4-metrics on $M$, $g_1=\Omega^2g_2$, can be chosen to  have ambient metrics  differing  by $\mathring{g}_1|_M(0)-\mathring{g}_2|_M(0)=g_1$.
\end{theorem}

This concludes the discussion of our ambient construction. In the next two sections, we move on to examine two important implications of the ambient geometric structures studied so far.

\section{Conformal structure and singularities}
Suppose that we start in step 1 of the ambient algorithm with a spacetime $(M,g_{\textrm{\textsc{in}}}(x^\mu))$ which in step 2 gives a conformally related 4-dimensional metric $g_4=\Omega^2g_{\textrm{\textsc{in}}}(x^\mu)$. Then it follows from the work we did in  previous sections that these two metrics have ambient metrics with the following property. To $g_4$, there is an ambient 5-metric $g_+$ satisfying the 5-dimensional Einstein equations with a fluid source such that its conformal infinity is described by the 4-dimensional metric $\mathring{g}|_M$ that belongs to the conformal  class $[g_4]$  and is a constant $w$-rescaling of $g_4$.
Similarly, we can construct the ambient metric of  $(M,g_{\textrm{\textsc{in}}}(x^\mu))$ (or of any other metric in the conformal geometry $[g_4]$) which will satisfy the 5-dimensional Einstein equations with a fluid source and  the metric  $\mathring{g}_{\textrm{\textsc{in}}}|_M$, will be a different $w$-rescaling of $g_4$. The difference of the two   ambient metrics $\mathring{g}_{\textrm{\textsc{in}}}|_M$ and $\mathring{g}_{4}|_M$ corresponding to the two conformally related boundary metrics $g_{\textrm{\textsc{in}}}$ and $g_4$ equals the initial metric $g_4$.

Therefore, since for a given $g\in [g_4]$, $\mathring{g}|_M$ is an honest metric we may imagine our 4-dimensional universe described as the conformal infinity of the ambient 5-metric $g_+$.
In this case, the properties of our 4-dimensional world would be dictated by the 5-dimensional ambient spacetime and the ambient metric $g_+$ satisfying the 5-dimensional Einstein-fluid equations. Then, various  fundamental cosmological questions acquire novel meanings:
Will there be any singularities in the 4-dimensional universe $(M,\mathring{g}|_M)$? Is cosmic censorship valid  on $(M,\mathring{g}|_M)$? What is the behaviour  of the various fields living on the ambient space at their  conformal infinity? What is the behaviour of entropy on $M$, and how can we distinguish initial from final singularities there? What is the relation of the properties of $(M,[g])$ to quantum gravity? Are there any  observational tests that point to the structure of $(M,\mathring{g}|_M)$?

For the ambient 5-spacetime $V$, these are all deep and difficult questions, and so the reader may wonder what we have gained with our construction. In general,   given a conformal structure on the boundary 4-spacetime $(M, [g_4])$ as we have done above, the ambient cosmologies will not of course have a form like the one we assumed above, namely, $a^2(w)g_4+dw^2$,  since we do not know the global solution to the 5-dimensional Einstein equations with fluid sources in the ambient spacetime (and even if we did, it would probably be exceedingly difficult to extract any information by elementary arguments).

However, suppose that we are not interested in what happens everywhere in the ambient spacetime $V=M\times\mathbb{R}$, but  only care about the physics on $M$. In this paper, we have constructed the notion of an ambient cosmology by starting from a conformal structure on $M$, and found that there is an ambient metric $g_+$ described in terms of formal power series solutions  locally in $V$ about $M$ of the 5-dimensional Einstein equations with fluid sources in $V$. Corresponding to the conformal structure on $M$, the ambient metric $g_+$ on $V$ has conformal infinity  that inherits the asymptotic properties  of a `nice' metric $g_4$ in the conformal $M$-geometry and is described as a homothetic multiple of $g_4$.

Therefore in this model, all physical properties of $M$  are captured and imprinted on it by its conformal structure $(M, [g_4])$ and the metric $\mathring{g}|_M$, which is in turn determined by the ambient metric formal series solutions about $M$ of the 5-dimensional Einstein equations with fluid sources. One asks: How do the asymptotic properties of the ambient metric $g_+$ and thermodynamic quantities such as the density $\rho_+$ of the 5-dimensional solutions affect the behaviour on $M$? For example, if $\rho_+$ diverges on points of $M$, or everywhere on $M\times\{0\}$, how will such a behaviour on $V$ affect that on $M$? As we have shown in previous sections, the form of the $M$-restriction of the ambient metric $\mathring{g}=\sigma(w)g_4$, for a $g\in[g_4]$, is determined by the smooth function $\sigma(w)$ and  the original 4-metric $g_4$, so that it reads, $\mathring{g}=(c_0+c_1w+\cdots)g_4$, with the constants determined by the 5-dimensional solutions of the Einstein equations with the fluid source. Hence, the possibly diverging behaviour of the 5-dimensional density $\rho_+$ at $M$, passes on to the metric $\mathring{g}|_M$ on $M$ only through the constants $c_0,c_1,\cdots$, in the sense that the different possible behaviours of $\rho$ would just readjust the values of these constants in a new $\mathring{g}|_M$.

We may therefore conclude that any singularities present in the metric $\mathring{g}|_M$  will be those remaining in $g_4$ after the conformal `cleaning' of $g_{\textrm{\textsc{in}}}(x^\mu)$ in step 2. No other new ones will arise from the ambient spacetime  $V$ following our construction. What are the constraints on $g_4$? The metric $g_4$ is seen to satisfy two conditions: The first is that it is constructed as in step 2 following the application of the conformal method \cite{pen86} in step 1, and the second is that it must comply with the basic condition Eq. (\ref{set1}). This can  give an incompatible constraint on $\mathring{g}|_M$ only when $g_4$ (and consequently, any homothetic multiple of it) becomes degenerate somewhere on $M$. This is discussed more fully in the  next section.

Compared with the standard situation described in general relativity as having the initial metric  $(M,g_{\textrm{\textsc{in}}}(x^\mu))$   satisfying the 4-dimensional Einstein equations with some sources, our situation described by the metric $\mathring{g}|_M$ induced on $M$ (and belonging in its conformal structure as above) constitutes a considerable improvement.
 The 4-metric $\mathring{g}|_M$ imposed on $M$ by the 5-dimensional ambient model discussed above inherits the properties of the `nice' metric $g_4$ (which in turn cannot be a solution of the same  4-dimensional Einstein equations satisfied by $g_{\textrm{\textsc{in}}}(x^\mu)$ because it is related conformally to the latter), while it does not receive any new singularities from the 5-dimensional ambient metric because it lies in its conformal infinity as described in this paper.

\section{Non-degeneracy and cosmic censorship}
 In this work we have shown that there is a relation between the conformal structure of the boundary geometry and the five-dimensional ambient metric construction.  The ambient metric leads to the asymptotic condition  (\ref{set1}), $\mathring{g}|_M=cg_4$, satisfied  by the  metrics  $g_4$ and $\mathring{g}|_M\in[g_4]$, and this imposes a compatibility condition on the two homothetically-related metrics. In particular, since $\mathring{g}|_M=\sigma g$ is by its construction non-degenerate on $M$, this means that $cg_4$ must also be a non-degenerate metric. We note that the latter is obtained by finding the conformal infinity of the ambient metric corresponding to the `nice' metric $g_4$, and then any metric $g$ in the conformal class of $g_4$ with have this form.  Thus, non-degeneracy of the $g_4$ metric in step two of the ambient algorithm, is an immediate implication of the ambient construction on the conformal structure of  $M$. (We note that metric reversal does not produce degeneracy on $g_4$.) Is it possible that $g_4$, in the process of step 2, develops singularities or other structures that make the ambient condition (\ref{set1}) an incompatible constraint for the ambient construction? This is a question  about the structure of the singular boundaries of the two metrics in (\ref{set1}), which we address below.

 We first note that a singular metric $g_4$ does not necessarily make the five-dimensional ambient metric $g_+$ (or the metric $\mathring{g}$) degenerate, unless it becomes null or timelike  somewhere. Therefore if a $g_4$ develops a singularity that is  spacelike everywhere, then $\mathring{g}$  will still be non-degenerate when restricted  on the boundary. (For example, such will be the case for the singular part of the horizon in non-static coordinates for the Schwarzschild black hole, while the null part is regular.)

 In general, non-degeneracy of the five-dimensional metrics $\mathring{g}$ or $g_+$ means that they do not become degenerate on important subspaces of the ambient spacetime, especially, this must be so at $y=0$, so we require that these metrics, when restricted on the boundary spacetime $M$, must not become degenerate. Since the signature of the five-dimensional metrics is $(1,3+1)$ (meaning one $-$ and four $+$'s), non-degeneracy implies that on the conformal boundary $M$ the restriction of the metric $g_+$ must retain its signature $(1,4)$. Since the boundary spacetime signature is $(1,3)$, we find that the boundary 3-space signature should not globally change from $+++$ to something else, that is to null $-++$,  or  to timelike $---$, because such a change would make the five-dimensional metric degenerate (in the sense that a null surface could then form in the ambient spacetime).

We propose that the choice of metric in the conformal class $[g_4]$ in step 2 of the ambient procedure of Section 2,  must be made such that it does not spoil the non-degeneracy of the $\mathring{g}$ metric when restricted along the boundary $M$. As discussed above, the only way then left for which the five-dimensional ambient metric will loose its non-degeneracy on $M$ is when a timelike or null hypersurface forms somewhere in  $\mathring{g}|_M$, that is when there are naked points at infinity on the boundary spacetime. This then would make the ambient cosmology  $\mathring{g}|_M$ degenerate, and  the difference $\hat{g}|_M$ in  (\ref{ac*}) will not make sense, a contradiction because as we showed this difference equals a rescaling of a well-defined spacetime metric. Therefore it seems that a choice must be made of those metrics $g_4$ in step 2 of the ambient procedure that respect cosmic censorship\footnote{The only other possibility is that the non-degeneracy of the ambient metric is spoiled when spherical symmetry is violated, for example when a suitable observer who falls into a black hole in an exact Kerr metric with $a<m$ (with $m,a$ constants, $m$ being the mass and $ma$ the angular momentum as measured from infinity), to whom the singularity is actually naked. However, there are reasons to believe  that such a situation inside the event horizon  is actually unstable, cf. \cite{pen}, p. 234, and \emph{generically} perturbed Kerr solutions will always develop a spacelike singularity in the neighborhood of its Cauchy horizon. A spacelike character of cosmological singularities is also supported by the BKL studies, cf. \cite{bkl} and refs. therein.}.

Conversely, the absence of naked singularities that follows from the validity of the asymptotic (in the sense of being valid on $\mathscr{I}_V$) condition (\ref{set1}) of the ambient cosmology has important implications.
For example, it follows that a naked singularity may not be the end product of the process of Hawking evaporation of a black hole through thermal radiation. In this case,  future null infinity will generically meet the vertical line coming out of the spacelike singularity of the black hole due to the evaporation (compare the Penrose diagrams in Figs. 3 and 5 of Ref. \cite{haw75}), thus allowing material from inside the  spacelike singularity to be seen by an observer sitting at infinity. This is sometimes interpreted, as is well-known, as a possible violation of cosmic censorship at the quantum level, a complete loss of predictability in a quantum treatment of black holes \cite{haw76}, and is also intimately related to the possible loss of information connected with the inevitable increase of entropy during this process \cite{bek73}.


\emph{Note added in proof:} After the completion of this work, we realized that there are certain analogies with some of the ideas of conformal cyclic cosmology \cite{pen2010}. The idea of masslessness near the singularities and a possible universal decay of all particle masses asymptotically  in \cite{pen2010}, may also be applicable here and associated with the crucial role played  by conformal (instead of Riemannian) geometry in the 4-dimensional conformal infinity of the 5-dimensional ambient metric.  However, in our work, we only have 5-dimensional Einstein equations with sources for the ambient metric. Another central characteristic implication of our work is that the existence of black holes and the validity of cosmic censorship in the 4-dimensional boundary spacetime seem to be intimately connected with the ambient construction taking place in the extra dimension, because they appear as properties of its conformal infinity. We believe it is an interesting question whether the approach developed here allows or even requires any cyclicity in the 4-dimensional conformal manifold.

\section*{Acknowledgements}
We thank Ioannis Bakas, Manolis Floratos and Kyriakos Papadodimas for useful discussions, and an anonymous referee for many useful and constructive comments which helped to present this work in a more comprehensive way.

\end{document}